\documentclass[twocolumn]{aastex61}

\usepackage{multirow}
\usepackage{color}

\received{February 13, 2019}
\accepted{April 3, 2019}

\shorttitle{Molecular Gas Properties in the Host Galaxy of GRB~080207}
\shortauthors{Hatsukade et al.}

\begin{document}

\title{Molecular Gas Properties in the Host Galaxy of GRB~080207}

\author{Bunyo Hatsukade}
\affiliation{Institute of Astronomy, Graduate School of Science, University of Tokyo, 2-21-1 Osawa, Mitaka, Tokyo 181-0015, Japan}
\email{hatsukade@ioa.s.u-tokyo.ac.jp}

\author{Tetsuya~Hashimoto}
\affiliation{National Tsing Hua University, No. 101, Section 2, Kuang-Fu Road, Hsinchu, 30013, Taiwan}

\author{Kotaro Kohno}
\affiliation{Institute of Astronomy, Graduate School of Science, University of Tokyo, 2-21-1 Osawa, Mitaka, Tokyo 181-0015, Japan}
\affiliation{Research Center for the Early Universe, The University of Tokyo, 7-3-1 Hongo, Bunkyo, Tokyo 113-0033, Japan}

\author{Kouichiro Nakanishi}
\affiliation{National Astronomical Observatory of Japan, 2-21-1 Osawa, Mitaka, Tokyo 181-8588, Japan}
\affiliation{SOKENDAI (The Graduate University for Advanced Studies), 2-21-1 Osawa, Mitaka, Tokyo 181-8588, Japan}

\author{Kouji Ohta}
\affiliation{Department of Astronomy, Kyoto University, Kyoto 606-8502, Japan}

\author{Yuu Niino}
\affiliation{Research Center for the Early Universe, The University of Tokyo, 7-3-1 Hongo, Bunkyo, Tokyo 113-0033, Japan}
\affiliation{Institute of Astronomy, University of Tokyo, 2-21-1 Osawa, Mitaka, Tokyo 181-0015, Japan}

\author{Yoichi Tamura}
\affiliation{Division of Particle and Astrophysical Science, Graduate School of Science, Nagoya University, Nagoya 464-8602, Japan}

\author{L. Viktor T{\'o}th}
\affiliation{Department of Astronomy of the E{\"o}tv{\"o}s Lor{\'a}nd University, P{\'a}zm{\'a}ny P{\'e}ter s{\'e}t{\'a}ny 1, 1117 Budapest, Hungary} 
\affiliation{Konkoly Observatory of the Hungarian Academy of Sciences, H-1121 Budapest, Konkoly Thege Mikl{\'o}s {\'u}t 15-17}


\begin{abstract}
We present the results of CO(1--0) and CO(4--3) observations of the host galaxy of a long-duration gamma-ray burst GRB~080207 at $z = 2.0858$ by using the Karl G. Jansky Very Large Array and the Atacama Large Millimeter/submillimeter Array. 
The host is detected in CO(1--0) and CO(4--3), becoming the first case for a GRB host with more than two CO transitions detected combined with CO(2--1) and CO(3--2) in the literature. 
Adopting a metallicity-dependent CO-to-H$_2$ conversion factor, we derive a molecular gas mass of $M_{\rm gas} = 8.7 \times 10^{10}$~$M_{\odot}$, which places the host in a sequence of normal star-forming galaxies in a $M_{\rm gas}$--star-formation rate (SFR) plane. 
A modified blackbody fit to the far-infrared--millimeter photometry results in a dust temperature of 37 K and a dust mass of $M_{\rm dust} = 1.5 \times 10^8$~$M_{\odot}$. 
The spatially-resolving CO(4--3) observations allow us to examine the kinematics of the host. 
The CO velocity field shows a clear rotation and is reproduced by a rotation-dominated disk model with a rotation velocity of 350 km~s$^{-1}$ and a half-light radius of 2.4 kpc. 
The CO spectral line energy distribution derived from the four CO transitions is similar to that of starburst galaxies, suggesting a high excitation condition. 
Comparison of molecular gas properties between the host and normal (main-sequence) galaxies at similar redshifts shows that they share common properties such as gas mass fraction, gas depletion timescale, gas-to-dust ratio, location in the $M_{\rm gas}$--SFR (or surface density) relation, and kinematics, suggesting that long-duration GRBs can occur in normal star-forming environments at $z \sim 2$.
\end{abstract}


\section{Introduction} \label{sec:introduction}

Long-duration gamma-ray bursts (GRBs) have been shown to be associated with the explosions of massive stars \citep[e.g.,][]{hjor03, stan03}. 
Due to the short lifetime of massive stars, GRBs are thought to trace galaxies with ongoing star formation. 
Because GRBs are bright enough to be observable in the cosmological distances \citep[e.g.,][]{tanv09, salv09}, they are expected to be a new tool to probe the star-forming activity in the distant universe \citep[e.g.,][]{tota97, wije98, kist09, robe12, tren12}. 
However, whether GRBs can be used as an unbiased tracer of star formation in the universe is still a subject of debate \citep[e.g.,][]{mich12, perl13, hunt14, grei15, perl15, perl16a, verg15, jape16}. 
In order to understand the environment where GRBs occur, it is important to understand the properties molecular gas, the fuel of star formation, in GRB host galaxies. 
While molecular hydrogen has been detected in absorption in the spectra of GRB afterglows \citep[e.g.,][]{proc09, kruh13, deli14, frii15}, the detected column density probes only one sight line in front of the GRB. 
Emission lines need to be observed in order to measure the molecular gas content in host galaxies. 
Searches for CO emission as a tracer of molecular gas have been conducted in several GRB host galaxies \citep{kohn05, endo07, hats07, hats11, stan11}, and five host galaxies have been detected so far \citep{hats14, stan15, mich16, arab18, mich18}. 
While earlier works suggest a deficiency of molecular gas in the host galaxies compared to their star formation rate (SFR) or stellar mass \citep{hats14, stan15}, recent studies show that the difference is not significant and GRB host galaxies have more diverse in molecular gas properties with an additional sample, appropriate CO-to-H$_2$ conversion factors ($\alpha_{\rm CO}$), or a careful choice of comparison sample \citep{arab18, mich18}.

One important issue for estimating the molecular gas mass is the uncertainty of CO line ratios. 
The CO luminosity of a ground rotational transition $J = 1$--0 is required for deriving the molecular gas mass by applying the conversion factor, and we need to assume a CO line ratio in the case of host galaxies with only higher $J$ CO lines. 
However, no GRB host galaxy has been detected with multiple CO transitions including the CO(1--0) line, and the CO line ratio for GRB host galaxies has not yet been obtained. 
Since the CO spectral line energy distribution (SLED) provides the molecular gas excitation condition whether GRB host galaxies have a nearly thermalized condition as in starburst galaxies or a low excitation condition as in the Milky Way, it is essential to obtain a sample of GRB host galaxies with multiple CO transitions.

In this paper, we present the results of CO(1--0) and CO(4--3) observations of the host galaxies of GRB~080207 with the Karl G. Jansky Very Large Array (VLA) and the Atacama Large Millimeter/submillimeter Array (ALMA). 
The arrangement of this paper is as follows. 
Section~\ref{sec:grb080207} outlines the host galaxy of GRB~080207. 
Section~\ref{sec:observations} describes the observations, data reduction, and results. 
In Section~\ref{sec:discussions}, we derived the molecular gas mass, dust mass, gas kinematics, and CO SLED, and compare them with other galaxies populations. 
We discuss the availability of GRBs as a tracer of star-formation activity in terms of molecular gas properties. 
Conclusions are presented in Section~\ref{sec:conclusions}. 
Throughout the paper, we adopt a cosmology with $H_0=70$ km s$^{-1}$ Mpc$^{-1}$, $\Omega_{\rm{M}}=0.3$, and $\Omega_{\Lambda}=0.7$. 
SFRs in this paper are converted to a \cite{chab03} IMF from a \cite{salp55} IMF by multiplying a factor of 0.61.

\section{The Host Galaxy of GRB~080207} \label{sec:grb080207}

The host galaxy of GRB~080207 has been well studied at various wavelengths. 
GRB~080207 is identified as a dark GRB due to the non-detection of its afterglow in the optical and NIR and the inferred X-ray-to-optical spectral slope \citep{hunt11, sven12, ross12}. 
The redshift of $z = 2.0858 \pm 0.0003$ is determined from the emission line in from the host galaxy \citep{kruh12}. 
The host galaxy is a red, massive, and actively star-forming galaxy with stellar mass of $M_* \sim 10^{11}$~$M_{\odot}$  and SFR of 50--300 $M_{\odot}$~yr$^{-1}$ \citep[e.g.,][]{perl13, hunt14, arab18}. 
The host galaxy has one of the highest stellar mass and inferred visual extinction ($A_V \gtrsim 2$ mag) among GRB hosts \citep[e.g.,][]{hunt11, hunt14, perl13, perl16b}. 
The stellar mass and SFR place the host galaxy on the massive end of the main sequence (MS) of star-forming galaxies at $z = 2$.
The detections at 24 $\mu$m with {\sl Spitzer}/MIPS \citep{hunt11, sven12}, 100 and 160 $\mu$m with {\sl Herschel}/PACS \citep{hunt14}, and 5 GHz with VLA \citep{pp13} suggest the existence of dust-obscured star formation. 
The radio-derived SFR of $846 \pm 124$ $M_{\odot}$~yr$^{-1}$ is larger than other estimates \citep{pp13}. 
While the radio observations were conducted 3.4 and 4.3 years after the GRB, the lack of early-time radio observations prevents constraining a long-lived radio afterglow, and its contamination to the detected radio emission from the host galaxy cannot be ruled out \citep{pp13}. 
The metallicity of the host galaxy is $12 + \log{\rm (O/H)} = 8.74 \pm 0.15$ \citep{kruh15}, which is derived with ${\rm N2}$ and ${\rm O3N2}$ diagnostics based on the calibrators of \cite{naga06} and \cite{maio08}. 
The host galaxy is on the mass-metallicity relation at $z \sim 2$ for the stellar mass \citep[e.g.,][]{wuyt14, stei14}.

Recently, \cite{arab18} detected the CO(3--2) line with the Plateau de Bure/NOrthern Extended Millimeter Array. 
They derived the molecular gas mass of $M_{\rm gas} = 1.1 \times 10^{11}$~$M_{\odot}$ and found that the gas mass fraction and the gas depletion timescale are comparable to star-forming galaxies at similar redshifts. 
\cite{mich18} reported the detection of the CO(2--1) line with the IRAM 30~m telescope and a derived molecular gas mass of $M_{\rm gas} = 2 \times 10^{11}$~$M_{\odot}$. 
The detection of the CO lines and the large gas reservoir make it the best target for studying the properties of molecular gas in detail.

\section{Observations and Results} \label{sec:observations}

\subsection{VLA CO(1--0)} \label{subsec:vla}

The VLA Ka-band observations (Project ID: 18A-103) were performed on Nov. 11--16, 2018 using 26 antennas. 
The array configuration was D and D$\rightarrow$C (transition from D to C) with a baseline length ranging from 40~m to 3.4~km. 
The WIDAR correlator was used with 8-bit samplers. 
We used two basebands with a 1 GHz bandwidth centered at 36.34 and 37.34 GHz, which provides a total bandwidth of 2 GHz centered at 36.8 GHz (0.81 cm). 
The redshifted CO(1--0) line was observed at 37.355 GHz. 
Bandpass and amplitude calibrations were done with 3C286, and phase calibrations were done with a nearby quasar J1415$+$1320. 
The total observing time and the on-source time is 7.4 and 3.7 hours, respectively.

The data were reduced with Common Astronomy Software Applications \citep[CASA;][]{mcmu07}. 
The maps were produced with \verb|tclean| task. 
The natural weighting was adopted, and the resultant synthesized beamsize is $1\farcs87 \times 1\farcs60$ (PA $= -0.45^{\circ}$) for  velocity-integrated CO(1--0) map and $1\farcs91 \times 1\farcs60$ (position angle PA $= -25^{\circ}$) for the continuum map. 
The continuum map was created with a total bandwidth of $\sim$1.9~GHz, excluding channels with emission line. 
The rms noise level is 7.5~$\mu$Jy~beam$^{-1}$ for the continuum map, and 83~$\mu$Jy~beam$^{-1}$ for the spectra with a velocity resolution of 100~km~s$^{-1}$.

\subsection{ALMA CO(4--3)} \label{subsec:alma}

ALMA observations were conducted in Apr. 29, 2016 for a Cycle 3 program (Project code: 2015.1.00939.S) as part of an ALMA CO search toward a sample of long-duration GRB host galaxies (B.~Hatsukade et al. in prep.). 
The redshifted CO(4--3) line was observed with Band 4. 
The correlator was used in the time domain mode with a bandwidth of 1875 MHz (488.28~kHz $\times$ 3840 channels). 
Four basebands were used, providing a total bandwidth of 7.5 GHz centered at 2.1 mm. 
The array configuration was C36-2/3 with the baseline lengths of 15.1--640.0 m. 
The number of available antenna was 41, and the on-source integration time is 41 min. 
Bandpass, flux, and phase calibrations were done with J1256$-$0547, Titan, and J1347+1217, respectively.

The data were reduced with CASA, and maps were processed with a \verb|tclean| task. 
While Briggs weighting with \verb|robust| parameter of 0.5 was adopted for CO maps, natural weighting was adopted for a continuum map to maximize the sensitivity. 
The continuum map was created with a total bandwidth of $\sim$7.5~GHz, excluding channels with emission line. 
Clean boxes were placed when a component with a peak signal-to-noise ratio (S/N) above 5 is identified, and \verb|CLEAN|ed down to a $2\sigma$ level. 
The synthesized beamsize is $0\farcs90 \times 0\farcs83$ (PA $= -1.0^{\circ}$) for a velocity-integrated CO map, and $1\farcs21 \times 1\farcs03$ (PA $= -32^{\circ}$) for the continuum map. 
The rms noise level is 12.9~$\mu$Jy~beam$^{-1}$ for the continuum map, and 0.42~mJy~beam$^{-1}$ for a spectra with a velocity resolution of 20~km~s$^{-1}$.

\begin{figure}
\begin{center}
\includegraphics[width=\linewidth]{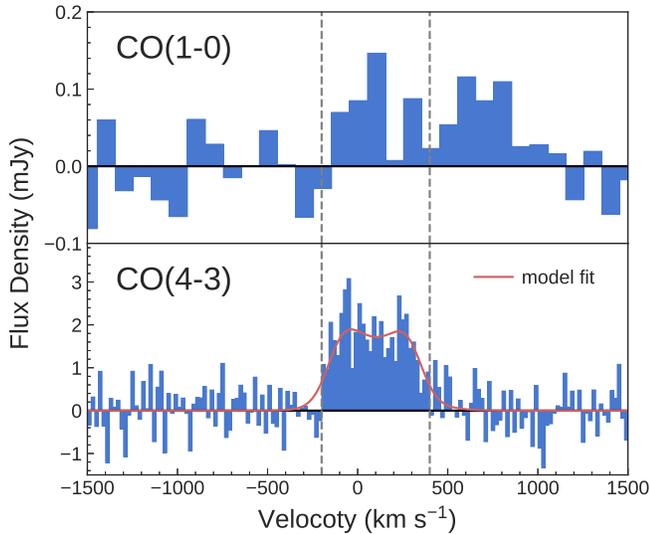}
\end{center}
\caption{
Line profiles of CO(1--0) (top) and CO(4--3) (bottom) of the GRB080207 host galaxy. 
Continuum emission is subtracted. 
The velocity resolution is 100 km~s$^{-1}$ and 20 km~s$^{-1}$ for CO(1--0) and CO(4--3), respectively. 
The horizontal axis is the velocity relative to $z = 2.0858$. 
The line flux is measured with a $1''$-radius aperture. 
The vertical dashed lines represent the velocity range used for measuring the flux. 
The red curve in the bottom panel shows the result of model fit with GalPaK$^{\rm 3D}$ (see Section~\ref{sec:kinematics}). 
}
\label{fig:spectra}
\end{figure}

\begin{figure*}[t]
\begin{center}
\includegraphics[width=.9\linewidth]{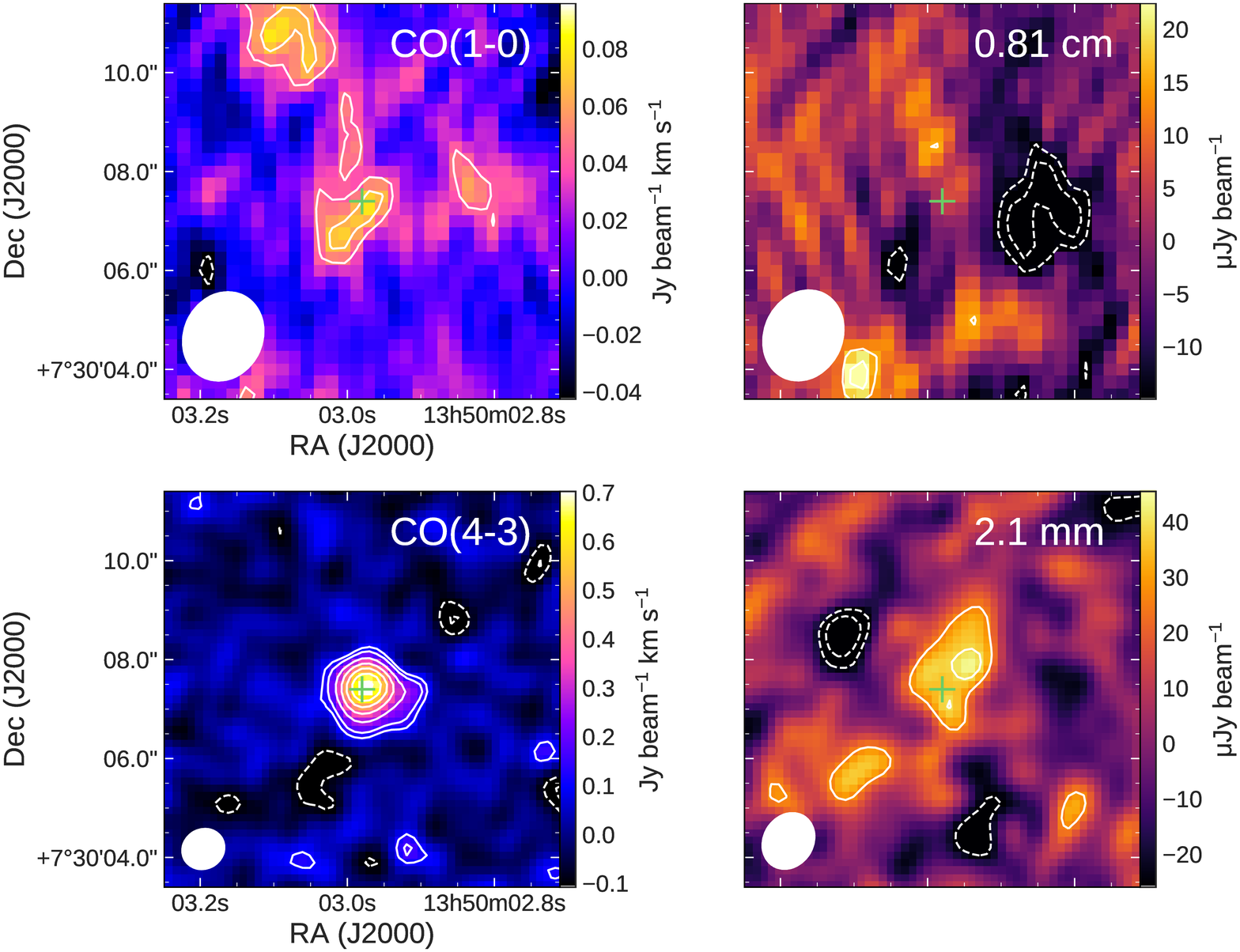} \\
\end{center}
\caption{
Maps of CO(1--0) integrated intensity (top left), 
0.81 cm continuum (top right), 
CO(4--3) integrated intensity (bottom left), 
and 2.1 mm continuum (bottom right). 
The same velocity range ($-200$ to 400 km~s$^{-1}$) is used creating the CO maps.
The images are $10'' \times 10''$ and the contours are $\pm$$2\sigma$, $\pm$$3\sigma$, and a $2\sigma$ step subsequently. 
Cross represents the GRB afterglow position. 
The synthesized beam size is shown in the lower left corners. 
}
\label{fig:map}
\end{figure*}

\subsection{Results} \label{sec:results}

We detected CO(1--0) and CO(4--3) emission at the position of the host galaxy and at the expected frequency for $z = 2.0858$ (Figure~\ref{fig:spectra} and \ref{fig:map}). 
The CO(4--3) spectrum shows a full width at zero intensity of $\sim$600 km~s$^{-1}$ with a range from $-200$ to 400 km~s$^{-1}$, and a full width at half maximum (FWHM) of $\sim$450 km~s$^{-1}$. 
This is consistent with the spectrum of CO(2--1) \citep{mich18} but is wider than is reported for CO(3--2) \citep[FWHM = $191 \pm 35$ km~s$^{-1}$;][]{arab18}.

The line profile of CO(1--0) appears to be somewhat different from CO(4--3), although the S/N is lower than that of CO(4--3). 
While the spectral feature at $\sim$300 km~s$^{-1}$ corresponds to the second peak of the CO(4--3) spectrum, the component at $\sim$700 km s$^{-1}$ is not seen in CO(4--3). 
The $\sim$300 km s$^{-1}$ and $\sim$700 km~s$^{-1}$ features are also marginally seen in the CO(3--2) spectrum of \cite{arab18}, and the CO(2--1) spectrum of \cite{mich18} has its peak at $\sim$300 km~s$^{-1}$. 
\cite{arab18} reported the presence of a broad component with a FWHM of 1160 km~s$^{-1}$ centered at 320 km~s$^{-1}$ in the H$\alpha$ emission, suggesting an outflow from the host galaxy or an interacting system. 
If we assume the $\sim$700 km~s$^{-1}$ component is a signature of outflow, the outflow mass is comparable to the main component of the host galaxy and is in the order of $10^{10}$--$10^{11}$~$M_{\odot}$. 
Such a massive molecular outflow is not seen in local ultra-luminous IR galaxies (ULIRGs) or Seyfert galaxies \citep[e.g.,][]{cico14, gonz17}. 
An interacting system is also a possible scenario. 
Signatures of interaction have been observed in GRB host galaxies through optical/near-IR imaging \citep[e.g.,][]{perl13} or absorption lines in afterglows \citep[e.g.,][]{thon13, wise17}. 
Neutral hydrogen observations of GRB host galaxies suggest a recent minor/major merger, which could induced star formation and the progenitor of a GRB formed in the star formation episode that took place in newly accreted gas \citep[e.g.,][]{mich15, arab15}. 
Since the $\sim$700 km~s$^{-1}$ component is not detected in other higher $J$ lines, the gas excitation condition would be low. 
Although the component could be a signature of an outflow or an interaction, the low significance (S/N $\sim 3.4$ in the velocity-integrated map) prevents further discussion. 
Note that no line or spurious can be found in the spectra of calibrators at the velocity range.

Since the CO spectra in the velocity ranging from $-200$ to $+400$ km~s$^{-1}$ share common features in all the CO transitions, we adopt the velocity range for measuring the line flux of CO(1--0) and CO(4--3) of the host galaxy. 
The peak S/N of the velocity-integrated intensity is 3.65 and 13.7 for CO(1--0) and CO(4--3), respectively. 
Since the significance for the CO(1--0) is not high, we conducted source extractions of positive peaks and negative peaks in the map to estimate the probability of spurious detection. 
The number of positive and negative peaks with S/N above 3.6 is 10 and 1, respectively, and the fraction of negative to positive peaks is 0.1. 
Given that the position of the CO(1--0) emission agrees with the host galaxy and the CO(4--3) emission within errors, we can consider the CO(1--0) originates from the host galaxy. 
The CO(1--0) emission is not spatially resolved and we adopt the peak value as a total flux. 
The CO(4--3) emission is spatially resolved with a deconvolved source size of $(0\farcs7 \pm 0\farcs2) \times (0\farcs5 \pm 0\farcs2)$. 
The CO line luminosities are 
$L'_{\rm CO}$(1--0) $= (1.7 \pm 0.5) \times 10^{10}$ (K~km~s$^{-1}$~pc$^2$) 
and $L'_{\rm CO}$(4--3) $= (1.5 \pm 0.1) \times 10^{10}$ (K~km~s$^{-1}$~pc$^2$).

Continuum emission at 0.81 cm is not detected with VLA, giving a 3$\sigma$ upper limit of $S_{\rm 0.81cm} < 22$ $\mu$Jy~beam$^{-1}$. 
Continuum emission at 2.1 mm is marginally detected with a peak S/N of 3.4. 
The emission is spatially resolved by the beam and the integrated flux measured with \verb|imfit| is $S_{\rm 2.1mm} = 120 \pm 20$ $\mu$Jy. 
The observation results are summarized in Table~\ref{tab:results}.

\begin{figure}[t]
\begin{center}
\includegraphics[width=\linewidth]{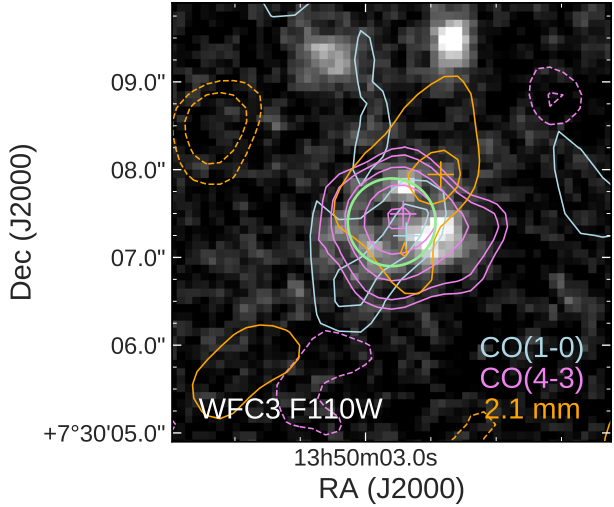} \\
\end{center}
\caption{
Contours of CO(1--0), CO(4--3), and 2.1 mm continuum overlaid on {\sl HST}/WFC3 F110W image. 
The image is $5'' \times 5''$ and the contours are $-3, -2, +2, +3, +5, +9, +13\sigma$. 
Crosses represent the peak positions. 
The 0\farcs5 error circle of the GRB afterglow position obtained from X-ray observations \citep{sven12} is presented by the green circle. 
}
\label{fig:hst}
\end{figure}

Figure~\ref{fig:hst} shows the contours of the CO(1--0) line, the CO(4--3) line, and the 2.1~mm dust continuum emission overlaid on the image of {\sl HST}/Wide Field Camera-3 (WFC3) with the F110W filter. 
The astrometry of image is calibrated by the source catalog of Gaia Data Release 1 \citep{gaia16}. 
The peaks of the CO(1--0) and CO(4--3) lines are located at the center of the optical components and consistent with the GRB afterglow position within errors. 
The peak of the dust continuum appears to be in-between the brighter north and west optical components, although the positional uncertainty is large ($\sim$$0.4''$). 
The discrepancy in the position between optical and CO/dust suggests the obscured star formation in the host galaxy.

\begin{table*}[t]
\begin{center}
\caption{Results.} \label{tab:results}
\begin{tabular}{ccccc@{\extracolsep{10pt}}cc@{}}
\hline\hline
Observations & \multicolumn{4}{c}{CO} & \multicolumn{2}{c}{Continuum} \\ \cline{2-5} \cline{6-7}
 & $J$ & $S_{\rm CO}\Delta v$$^*$ & $L'_{\rm CO}$ & Deconvolved Size$^\dagger$ & $S_{\rm cont}$ & Deconvolved Size$^\dagger$ \\
             &    & (Jy km s$^{-1}$)         & (K km s$^{-1}$~pc$^2$) & ($''$) & ($\mu$Jy) & ($''$) \\
\hline
VLA  & 1--0 & $0.078 \pm 0.020$$^\ddagger$ & $(1.7 \pm 0.5) \times 10^{10}$ & $\cdots$ & $<$22 (3$\sigma$)& $\cdots$ \\
ALMA & 4--3 & $1.1 \pm 0.1$$^\dagger$ & $(1.5 \pm 0.1) \times 10^{10}$ & $(0.7\pm0.2)\times(0.5\pm0.2)$ & $120 \pm 20$$^\dagger$ & $(2.1\pm0.5)\times(1.0\pm0.3)$ \\
\hline
\multicolumn{7}{l}{\parbox{150mm}{
$^*$ Velocity-integrated CO intensity derived by integrating the velocity from $-200$ to $+400$ km s$^{-1}$. \\
$^\dagger$ Derived from 2D Gaussian fit with CASA {\sf imfit}. \\
$^\ddagger$ Peak intensity. 
}}
\end{tabular}
\end{center}
\end{table*}

\section{Discussion} \label{sec:discussions}

\subsection{Molecular Gas Mass} \label{sec:gas}

We derive the molecular gas mass based on the CO(1--0) emission.
The CO luminosity is calculated as follows \citep{solo05}, 
\begin{eqnarray}
L'_{\rm CO} = 3.25 \times 10^7 S_{\rm CO}\Delta v \nu_{\rm obs}^{-2} D_{\rm L}^2 (1+z)^{-3}, 
\end{eqnarray}
where $L'_{\rm CO}$ is in K km s$^{-1}$~pc$^2$, $S_{\rm CO}\Delta v$ is the velocity-integrated intensity in Jy~km~s$^{-1}$, $\nu_{\rm obs}$ is the observed line frequency in GHz, and $D_{\rm L}$ is the luminosity distance in Mpc.
The molecular gas mass is derived from 
\begin{eqnarray}
M_{\rm gas} = \alpha_{\rm CO} L'_{\rm CO(1-0)}, 
\end{eqnarray}
where the CO-to-H$_2$ conversion factor $\alpha_{\rm CO}$ includes the contribution of helium mass. 
The conversion factor is thought to be dependent on gas phase metallicity, increasing $\alpha_{\rm CO}$ with decreasing metallicity \citep[e.g.,][]{wils95, arim96, kenn12, bola13}. 
We adopt the relation between metallicity and $\alpha_{\rm CO}$ of \cite{genz12} and \cite{genz15}, where they derived an empirical relation by using local and high-redshift sample. 
To apply the relation, we convert the metallicity reported in \cite{kruh15} to the calibration of \cite{pett04} by using the metallicity conversion of \cite{kewl08}.
The metallicity of the host galaxy corresponds to a metallicity-dependent conversion factor of $\alpha_{\rm CO} = 5.1$ $M_{\odot}$~(K~km~s$^{-1}$~pc$^2$)$^{-1}$. 
If we adopt a Galactic conversion factor of $\alpha_{\rm CO} = 4.3$ $M_{\odot}$~(K~km~s$^{-1}$~pc$^2$)$^{-1}$ \citep{bola13}, a derived molecular gas mass would be smaller by a factor of 1.2. 
The molecular gas mass based on the CO(1--0) luminosity is estimated to be $M_{\rm gas} = (8.7 \pm 2.4) \times 10^{10}$~$M_{\odot}$, which is 1.5--2 times smaller than those derived from the CO(2--1) and CO(3--2) observations \citep{mich18, arab18}. 
The difference can be explained by the assumed CO line ratios in the literature: 
the brightness temperature line ratios CO(2--1)/CO(1--0) ($r_{21}$) of 0.5 in \cite{mich18} and CO(3--2)/CO(1--0) ($r_{31}$) of 0.6 in \cite{arab18}, while the derived ratios in this study are close to unity (see Sec.~\ref{sec:sled}). 
The host galaxy has molecular gas reservoir more than an order of magnitude compared to the other CO-detected GRB host galaxies of GRB~980425 at $z = 0.0085$, GRB~051022 at $z = 0.809$, GRB GRB~080517 at $z = 0.089$, and GRB~111005A at $z = 0.01326$ \citep{hats14, stan15, mich16, arab18, mich18}.

\begin{figure}[t]
\begin{center}
\includegraphics[width=\linewidth]{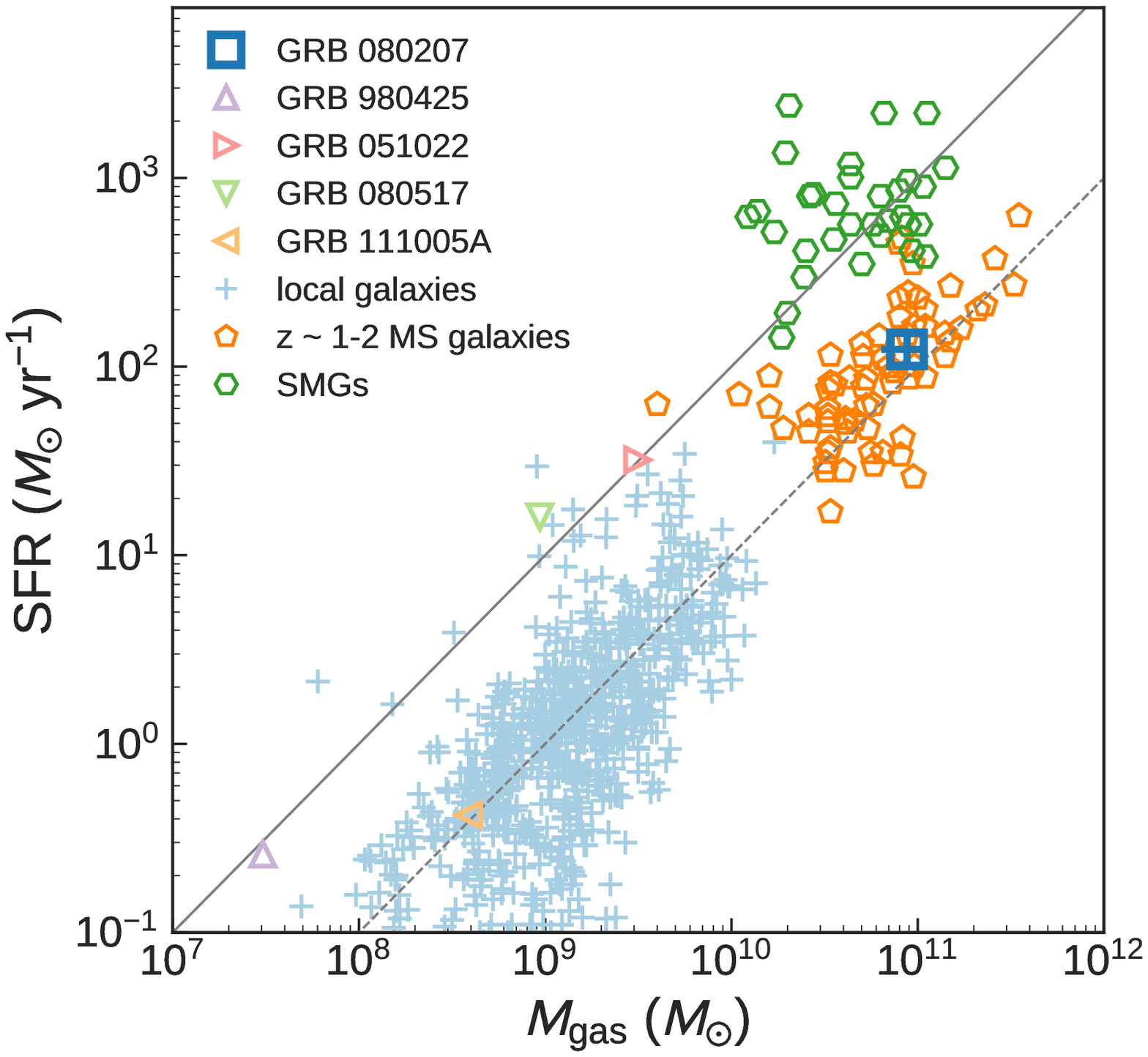}
\end{center}
\caption{
Comparison of molecular gas mass and SFR. 
We plot the CO-detected GRB host galaxies of GRB with metallicity-dependent conversion factor \citep{hats14, stan15, mich18}, 
local galaxies \citep{sain11, sain17, both14}, 
$z \sim 1$--2 MS galaxies \citep{tacc13}, 
and SMGs \citep{both13}. 
The solid and dashed lines represent gas depletion times of 0.1 and 1 Gyr, respectively. 
}
\label{fig:mgas-sfr}
\end{figure}

The derived molecular gas mass is compared with stellar mass and SFR. 
We adopt the results of SED fitting ($M_* = 1.7 \times 10^{11}$~$M_{\odot}$ and SFR $= 1.2 \times 10^2$~$M_{\odot}$~yr$^{-1}$) by T.~Hashimoto et al. (in prep.), where they include newly-obtained ALMA photometry in addition to UV--IR data. 
The molecular gas mass fraction defined as $M_{\rm gas}/(M_{\rm gas} + M_*)$ is 0.34, which is comparable to those of $z \sim 1$--2 MS galaxies with the similar stellar mass \citep{tacc10, tacc13, sarg14, seko16a}. 
The star-formation efficiency (SFR$/M_{\rm gas}$) and the gas depletion timescale ($M_{\rm gas}/$SFR) are estimate to be 1.4 Gyr$^{-1}$ and 0.71 Gyr, respectively. 
Figure~\ref{fig:mgas-sfr} shows the comparison between molecular gas mass and SFR. 
\cite{dadd10} presented two different star-forming regimes for disk-like normal star-forming galaxies and starburst galaxies in the $M_{\rm gas}$--SFR (or surface density) plane, which are applied for both local and high-$z$ galaxies. 
The GRB~080207 host galaxy is located in the similar region for $z \sim 1$--2 MS galaxies and on the sequence of disk-like normal star-forming galaxies.

\begin{table*}[t]
\begin{center}
\caption{Derived Properties.} \label{tab:properties}
\begin{tabular}{cccccccc}
\hline\hline
$M_{\rm gas}$ & $\Sigma_{\rm gas}$ & $f_{\rm gas}$ & SFE & $\tau_{\rm depl}$ & $\tau_{\rm dyn}$ & $M_{\rm dust}$ & $M_{\rm gas}/M_{\rm dust}$ \\
($M_{\odot}$) & ($M_{\odot}$~pc$^{-2}$) &          & (Gyr$^{-1}$) & (Gyr)    & (Myr)            & ($M_{\odot}$)  &     \\
(1)           & (2)                & (3)           & (4) & (5)               & (6)              & (7)            & (8) \\
\hline
$(8.7\pm2.4)\times 10^{10}$ & $2.4 \times 10^3$ & 0.34 & 1.4 & 0.71 & 42 & $1.5 \times 10^8$ & 560 \\
\hline
\multicolumn{8}{l}{\parbox{120mm}{
(1) Molecular gas mass derived with $\alpha_{\rm CO} = 5.1$ $M_{\odot}$~(K~km~s$^{-1}$~pc$^2$)$^{-1}$. 
(2) Molecular gas surface density within the CO half-light radius ($0.5 M_{\rm gas}/\pi R_{\rm 1/2,CO}^2$). 
(3) Molecular gas mass fraction ($M_{\rm gas}/(M_{\rm gas} + M_*)$). 
(4) Star-formation efficiency (SFR$/M_{\rm gas}$). 
(5) Gas depletion timescale ($M_{\rm gas}/$SFR). 
(6) Dynamical timescale at the CO half-light radius ($2\pi R_{\rm 1/2,CO}/v_{\rm rot}$).
(7) Dust mass derived form the modified blackbody fit. 
(8) Molecular gas mass to dust mass ratio. 
}}
\end{tabular}
\end{center}
\end{table*}

\subsection{Dust Properties} \label{sec:dust}

By using the far-infrared (FIR)--millimeter photometry obtained in this study and literature, we derived the FIR luminosity ($L_{\rm FIR}$) and dust mass ($M_{\rm dust}$) following \cite{debr03} as 
\begin{eqnarray}
& & L_{\rm FIR} = 4\pi M_{\rm dust} \int_0^{\infty} \kappa_d(\nu_{\rm{rest}})B(\nu_{\rm{rest}}, T_{\rm dust}) d\nu, \\
& & M_{\rm dust} = \frac{S_{\rm{obs}}D_L^2}{(1+z)\kappa_d(\nu_{\rm{rest}})B(\nu_{\rm{rest}}, T_{\rm dust})},
\end{eqnarray}
where $\kappa_d(\nu_{\rm{rest}})$ is the dust mass absorption coefficient, $\nu_{\rm{rest}}$ is the rest-frame frequency, $T_{\rm dust}$ is the dust temperature, $B(\nu_{\rm{rest}}, T_{\rm dust})$ is the Planck function, and $S_{\rm{obs}}$ is the observed flux density.
We assume that the absorption coefficient varies as $\kappa_d \propto \nu^{\beta}$ and adopt $\kappa_d(125\ \mu m) = 2.64 \pm 0.29$~m$^2$~kg$^{-1}$ \citep{dunn03}. 
To derive the dust temperature and emissivity index, we fit a modified black body to the photometry at 160~$\mu$m obtained with {\sl Herschel}/PACS \citep{hunt14}, 480~$\mu$m and 650~$\mu$m with ALMA Band 8 and 9 (T.~Hashimoto et al. in prep.), and our 2.1~mm with ALMA Band 4. 
The best-fit values are $T_{\rm dust} = 37 \pm 4$~K and $\beta = 1.5 \pm 0.3$. 
The dust temperature is comparable to the typical value of MS galaxies at $z \sim 2$ with a similar stellar mass \citep[e.g.,][]{magn14, genz15, schr18}. 
\cite{hunt14} estimated a 1.6 times higher dust temperature of 61.3~K for the host galaxy by fitting the SED with {\sc grasil} \citep{silv98}. 
A possible cause of the difference is that, in addition to the different approaches, they used only {\sl Herschel}/PACS data at FIR--millimeter wavelengths while we added new ALMA data points in the Rayleigh-Jeans regime at rest-frame wavelengths of 155, 205, and 680~$\mu$m. 
The derived dust temperature of $\sim$37~K is lower among the sample of FIR--submillimeter detected GRB host galaxies. 
\cite{mich08} studied four submillimeter-detected GRB hosts and derived the dust temperature of $T_{\rm dust} = 44$--51~K, and 
\cite{hunt14} derived dust temperatures of 17 GRB hosts with {\sl Herschel} observations, which range from 21 to 132~K with an average temperature of 48~K. 
The low dust temperature derived in this study is against the hypothesis that GRB host galaxies have warmer dust than in local ULIRGs and SMGs \citep{prid06, mich08, hunt14}.

The derived dust mass and FIR luminosity are $M_{\rm dust} = 1.5 \times 10^8$~$M_{\odot}$ and $L_{\rm FIR} = 1.4 \times 10^{12}$~$L_{\odot}$ (with 30\% error from 1$\sigma$ of the map), respectively. 
SFR is $= 150$~$M_{\odot}$~yr$^{-1}$ derived from SFR $= 1.72 \times 10^{-10} L_{\rm FIR}$ \citep{kenn98a} and scaled to \cite{chab03} IMF. 
The SFR is consistent with the results of SED fitting by Hashimoto et al. (in prep.). 
The molecular gas-to-dust mass ratio is $M_{\rm gas}/M_{\rm dust} \sim 560$, which is comparable to those of $z \sim 1.4$ MS galaxies with the similar stellar mass and metallicity \citep{seko16a, seko16b}.

\subsection{Kinematics} \label{sec:kinematics}

In order to study the kinematics and morphological parameters, we fit the spatially resolved CO(4--3) data with GalPaK$^{\rm 3D}$ \citep{bouc15}\footnote{http://galpak.irap.omp.eu/}, which is a code designed to extract the intrinsic galaxy parameters and kinematics directly from three-dimensional data cubes. 
The algorithm directly compares data-cubes with a disk parametric model with 10 free parameters:
coordinates of galaxy center ($x_c, y_c, z_c$), flux, half-light radius ($R_{1/2}$), inclination angle ($i$), position angle (PA), turnover radius of the rotation curve ($R_t$), de-projected maximum rotation velocity ($v_{\rm rot}$), and intrinsic velocity dispersion ($\sigma_0$). 
The algorithm uses a Markov Chain Monte Carlo (MCMC) approach and the knowledge of three-dimensional spread-functions. 
We adopt a disk model object with exponential flux profile, arctangent rotation profile, and thick disk dispersion.

\begin{figure}[t]
\begin{center}
\includegraphics[width=\linewidth]{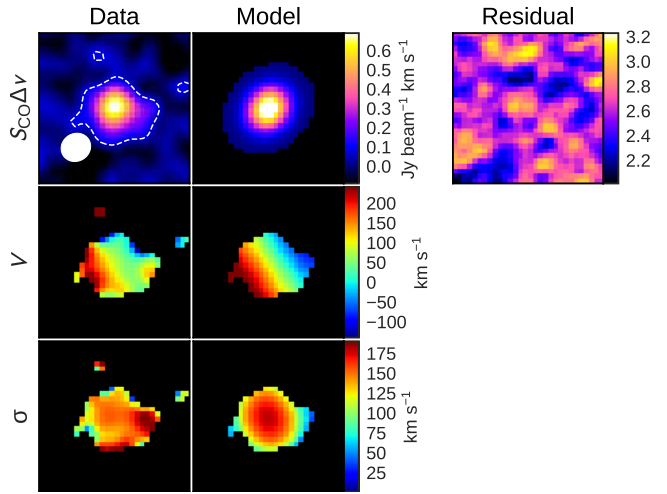}
\end{center}
\caption{
ALMA CO(4--3) images (1st row), the results of the GalPaK$^{\rm 3D}$ modeling (2nd row), and the residual (3rd row) generated from the residual cube and is normalized to the pixel noise. 
The 1st, 2nd, and 3rd column show the velocity-integrated intensity maps, intensity-weighted velocity field maps, and intensity-weighted velocity dispersion maps, respectively. 
The panel size is $4\farcs5 \times 4\farcs5$. 
The ALMA synthesized beamsize is shown in the lower left corner of the top-left panel. 
}
\label{fig:galpak}
\end{figure}

The procedure of model fitting converged with a reduced-$\chi^2$ of 1.12. 
The output images of the best-fit model are shown in Figure~\ref{fig:galpak} and the best-fit parameters are presented in Table~\ref{tab:galpak}. 
The uncertainty is the 95\% confidence interval calculated from the last 60\% of the MCMC chain for 15000 iterations.
Note that the output parameters have systematic errors of $\sim$10\% for different imaging parameters \citep{tada19}.
The CO line profile extracted from the best-fit model is presented in Figure~\ref{fig:spectra}. 
The best-fit model is a rotation-dominated disk with $Rr_{\rm 1/2,CO} = 2.4$~kpc, $v_{\rm rot} = 348$~km~s$^{-1}$, and $\sigma_0 = 63$~km~s$^{-1}$. 
The half-light radius is comparable to the deconvolved source size (Section~\ref{sec:results}), and a factor of 1.4 smaller than the optical half-light radius ($R_{\rm 1/2,opt}= 0\farcs4$ or 3.4~kpc) measured in {\sl HST}/F160W observations \citep{sven12}. 
The ratio of sizes between molecular gas and stellar component is within the range of $z \sim 1$--2 star-forming galaxies \citep{tacc13} measured with the CO(3--2) line. 
It is possible that the CO(4--3) emission traces denser, compact regions compared to lower $J$ lines. 
The ratio of rotational motion to velocity dispersion of $v_{\rm rot}/\sigma_0 = 5.5$ is comparable to those of $z \sim 1$--1.5 star-forming galaxies measured with CO lines \citep{tacc13} or $z \sim 2$ star-forming galaxies measured with near-infrared integral field spectroscopy \citep[e.g.,][]{fors09, fors18, genz11}.

The dynamical mass enclosed at a radius $R_{\rm 1/2,CO}$ and $R_{\rm 1/2,opt}$ calculated as $Rv_{\rm rot}^2/G$ is $\sim$$7 \times 10^{10}$~$M_{\odot}$ and $\sim$$9 \times 10^{10}$~$M_{\odot}$, respectively. 
The dynamical mass is comparable to the sum of the molecular gas mass and stellar mass, suggesting that the contribution of dark matter to the enclosed mass is not significant. 
This is consistent with the results on $z \sim 1$--2 MS galaxies \citep{fors09, bola15}.

\begin{table}[t]
\begin{center}
\caption{Best-fit Parameters of GalPaK$^{\rm 3D}$ Modeling} \label{tab:galpak}
\begin{tabular}{cc}
\hline\hline
Parameters & Results \\
\hline
$R_{\rm 1/2,CO}$ (kpc)      & $  2.39_{-0.14}^{+0.15}$ \\
$i$ (deg)                   & $ 38.74_{-1.41}^{+1.34}$ \\
PA (deg)                    & $127.35_{-3.53}^{+3.65}$ \\
$R_t$ (kpc)                 & $  0.037_{-0.025}^{+0.075}$ \\
$v_{\rm rot}$ (km~s$^{-1}$) & $347.59_{-8.08}^{+2.25}$ \\
$\sigma_0$ (km~s$^{-1}$)    & $ 63.44_{-5.95}^{+8.38}$ \\
\hline
\multicolumn{2}{c}{\parbox{80mm}{
Notes. - 
The uncertainty is the 95\% confidence interval.
}}
\end{tabular}
\end{center}
\end{table}

\subsection{Kennicutt-Schmidt Relation} \label{sec:ks-law}

\begin{figure}[t]
\begin{center}
\includegraphics[width=\linewidth]{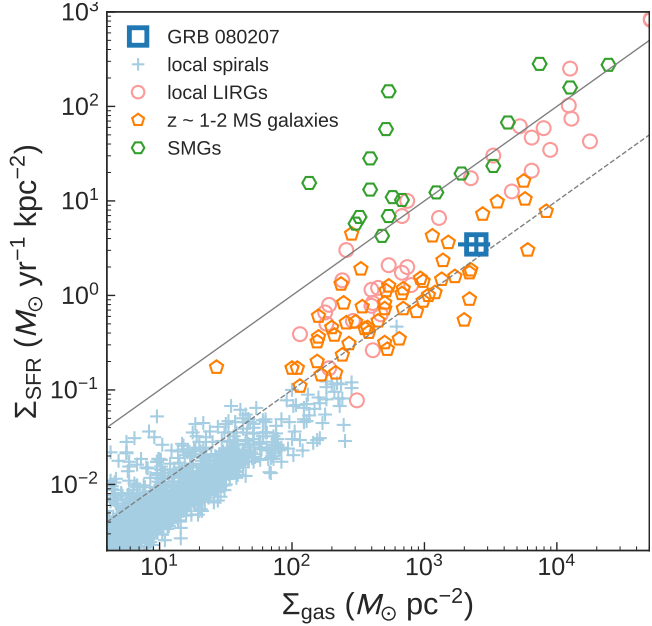}
\end{center}
\caption{
Comparison of molecular gas mass surface density and SFR surface density. 
For comparison, we plot other type of galaxies in the literature, where size measurements are available: 
local spirals \citep{kenn98b, bigi10}, 
local LIRGs \citep{kenn98b}, 
$z \sim 1$--2 MS galaxies \citep{tacc13}, 
and SMGs \citep{both10, genz10}. 
The solid and dashed lines represent gas depletion times of 0.1 and 1 Gyr, respectively. 
}
\label{fig:ks-law}
\end{figure}

There is a correlation between gas surface density ($\Sigma_{\rm gas}$) and SFR surface density ($\Sigma_{\rm SFR}$) known as the Kennicutt-Schmidt relation \citep{kenn98a, schm59}. 
Spatially-resolving observations of molecular gas are still limited and it is not well examined whether GRB host galaxies follow the the Kennicutt-Schmidt relation or not. 
\cite{arab18} studied CO-detected GRB host galaxies in the $\Sigma_{\rm gas}$--$\Sigma_{\rm SFR}$ plane by using the half-light radius of stellar emission for deriving the gas surface density. 
Our spatially-resolved molecular gas data allow us to directly estimate the molecular gas surface density. 
By using the CO half-light radius $R_{\rm 1/2,CO}$, the surface density of molecular gas and SFR is $\Sigma_{\rm gas} = 2.4 \times 10^3$~$M_{\odot}$~pc$^{-2}$ and $\Sigma_{\rm SFR} = 3.4$~$M_{\odot}$~yr$^{-1}$~kpc$^{-2}$, respectively.
Figure~\ref{fig:ks-law} shows the location of the GRB host in the $\Sigma_{\rm gas}$--$\Sigma_{\rm SFR}$ plane in comparison with other type of local and high-$z$ galaxies in the literature, where molecular gas and SFR surface densities are available. 
As in the case of $M_{\rm gas}$--$M_{\rm SFR}$ relation, the host galaxy is on the sequence of normal star-forming galaxies, indicating that the star-formation efficiency or gas deletion timescale is comparable to those of normal star-forming galaxies. 
This is consistent with the finding of \cite{arab18}, although they derived the molecular gas mass of the host galaxy from CO(3--2) observations and adopted $R_{\rm 1/2,opt}$ to estimate $\Sigma_{\rm gas}$.

\begin{figure}[t]
\begin{center}
\includegraphics[width=\linewidth]{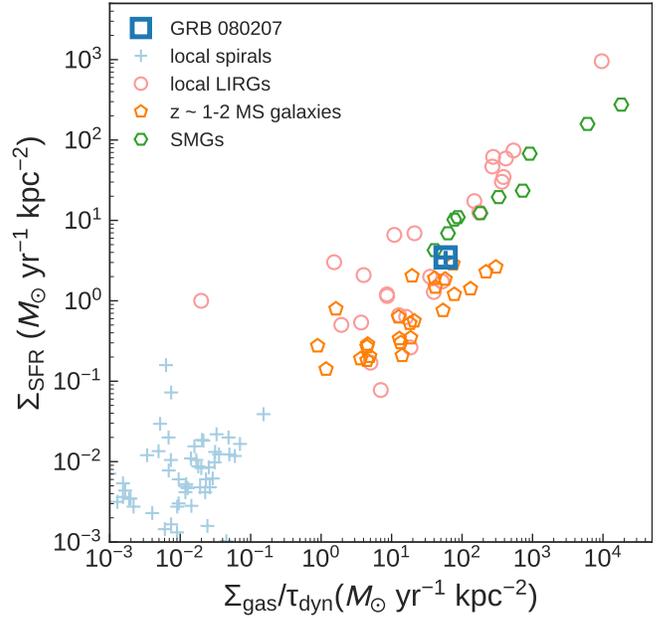}
\end{center}
\caption{
Relation between SFR surface density and the ratio of gas density to dynamical timescale. 
We plot other type of galaxies in the literature, where size measurements and dynamical timescale are available: 
local spirals and local LIRGs are from \citep{kenn98b}, and $z \sim 1$--2 MS galaxies and SMGs are from \citep{genz10}. 
}
\label{fig:sigmaGas-tdyn-sigmaSFR}
\end{figure}

\cite{kenn98a,kenn98b} showed that SFR surface density scales with the ratio of gas surface density to dynamical timescale in local spiral and starburst galaxies, and the relation is found to be valid for high-$z$ MS galaxies and SMGs \citep{dadd10, genz10}. 
The dynamical timescale of the GRB~080207 host galaxy at the CO half-light radius is 42~Myr calculated as $\tau_{\rm dyn} = 2\pi R_{\rm 1/2,CO}/v_{\rm rot}$. 
Figure~\ref{fig:sigmaGas-tdyn-sigmaSFR} shows that the host galaxy is on the general star-formation law for local and high-$z$ galaxies.

\subsection{CO Line Ratios} \label{sec:sled}

\begin{figure}[t]
\begin{center}
\includegraphics[width=\linewidth]{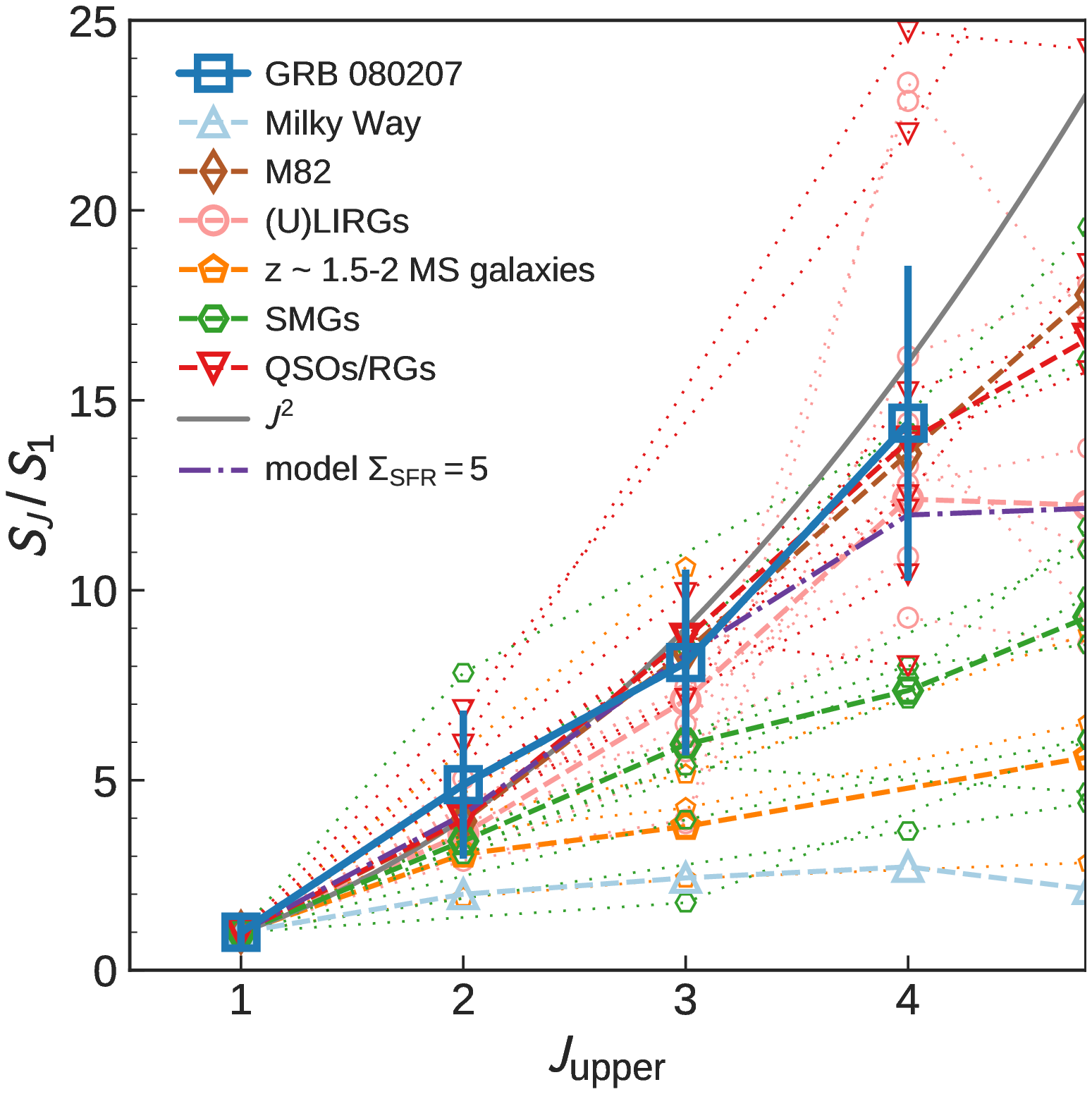}
\end{center}
\caption{
CO SLEDs normalized to the CO(1--0) line for the GRB~080207 host galaxy and other type of galaxies. 
The values for Milky Way, M~82, SMGs, and QSOs/radio galaxies (RGs) are taken from the compiled data in \cite{cari13}, (U)LIRGs are from \cite{papa12}, and $z \sim 1.5$ MS galaxies are from \cite{dadd15} and \cite{bola15}. 
We plot the (U)LIRGs, SMGs, and QSOs/RGs with CO detection up to $J \geq 4$. 
The average values are shown in bold dashed lines. 
The gray solid line represents constant brightness temperature ($J^2$). 
The gray dot-dashed line represents the model of \cite{nara14} with $\Sigma_{\rm SFR} = 5$~$M_{\odot}$~yr$^{-1}$~kpc$^{-2}$. 
}
\label{fig:sled}
\end{figure}

The ratios of CO line luminosities in GRB hosts galaxies have been unexplored due to the limited number of CO observations and the lack of detections of multiple CO transitions. 
Since the GRB~080207 host was detected with CO(2--1) and CO(3--2) in \cite{mich18} and \cite{arab15}, the detections of CO(1--0) and CO(4--3) in this study allow us to study the CO excitation condition with the upper $J$ level of from 1 to 4. 
The derived brightness temperature line ratios are summarized in Table~\ref{tab:line_ratio}. 
The line ratios of the GRB~080207 host show that molecular gas excitation is high and close to unity up to $J = 4$. 
In Figure~\ref{fig:sled}, we compare the CO SLEDs of the GRB~080207 host galaxy and other type of galaxies. 
The line ratio of the GRB~080207 host galaxy deviates from Milky Way and is similar to those of local starburst M~82, local (U)LIRGs, and QSOs/radio galaxies, suggesting the presence of dense and warm gas \citep[e.g.,][]{weis07,cari13}. 
While the average SLEDs of the MS galaxies at $z \sim 1.5$ and SMGs have a subthermal excitation, they show a diverse range of excitation conditions \citep{dann09, arav14, dadd15, bola15, both13}. 
\cite{bola15} reported that two $z \sim 2$ MS galaxies, which have a similar stellar mass and SFR to the GRB~080207 host galaxy, have high CO excitation conditions with $r_{31} \sim 0.9$--1.1, which is comparable to our results. 
There are many possible drivers for for high excitation conditions. 
\cite{dadd15} found that the ratio of CO(5--4) to lower-$J$ CO emission is well correlated with the average intensity of the radiation field and with $\Sigma_{\rm SFR}$ among local spirals, local ULIRGs, and $z \sim 1.5$ MS galaxies. 
This is consistent with the model of \cite{nara14}. 
They combined hydrodynamic simulations of disk galaxies and galaxy mergers with molecular line radiative transfer calculations, motivated to examine the origin of a broad range of observed CO SLEDs in local and high-redshift galaxies. 
They found that CO SLEDs are well correlated with the galaxy-wide SFR surface density, which is attributed to the increase of gas density and temperature. 
As shown in Figure~\ref{fig:sled}, we find that their model for galaxies with $\Sigma_{\rm SFR} \sim 5$~$M_{\odot}$~yr$^{-1}$~kpc$^{-2}$ agrees with our results.

It might be also possible that the high CO excitation condition is attributed to a strong radiation by an AGN. 
While no strong evidence for an AGN has been reported, the optical emission-line diagnostic with the Baldwin-Phillips-Terlevich (BPT) diagram \citep{bald81} studied in \cite{kruh15} shows that the host galaxy is located in a starburst-AGN composite region. 
The significantly high radio-derived SFR (850 $M_{\odot}$~yr$^{-1}$) derived in \cite{pp13} compared to other methods could be explained by the contribution of an AGN to the observed radio flux. 
Since peaks in CO SLED for AGNs are $J \gtrsim 8$--10 \citep[e.g.,][]{meij06}, where high ISM excitation occurs in X-ray dissociation regions, observations of higher $J$ CO lines are important to diagnose the presence of an AGN.

\begin{table}[t]
\begin{center}
\caption{CO Brightness Temperature Line Ratio} \label{tab:line_ratio}
\begin{tabular}{cc}
\hline\hline
Line & Ratio \\
\hline
$r_{21}$ & $1.22 \pm 1.04$ \\
$r_{32}$ & $0.90 \pm 0.46$ \\
$r_{43}$ & $0.90 \pm 0.26$ \\
\hline
\multicolumn{2}{c}{\parbox{35mm}{
Notes. - 
The error is 1$\sigma$. 
}}
\end{tabular}
\end{center}
\end{table}

\subsection{Implication for GRBs as a tracer of SFR history} \label{sec:implication}

There is ongoing debate as to whether GRBs occur in normal star-forming galaxies or occur under particular circumstances, and whether they can be used as an unbiased tool to trace star formation in the universe. 
It is suggested that GRBs prefer low-metallicity environments with a threshold of 0.3--0.7 $Z_{\odot}$ \citep[e.g.,][]{woos06, grah13, verg15, perl16b, verg17}. 
The metallicity effect on the GRB rate is modest at $z \gtrsim 2$, where the mean metallicity decreases, and GRBs can be unbiased tracers of star formation \citep[e.g.,][]{fynb06}. 
In addition to the metallicity effect, the reason for the smaller number of GRBs in massive galaxies \citep[e.g.,][]{perl13} can be attributed by dust obscuration. 
GRBs are usually followed by afterglow, which has been a major probe to identify the host galaxy and to study the interstellar medium. 
A significant fraction \citep[$\sim$25--40\%; e.g.,][]{fynb09, grei11} of GRB afterglows is optically dark compared with what is expected from X-ray afterglows, which are know as dark GRBs \citep{jako04, vand09}, preventing follow-up observations of host galaxies. 
A possible explanation of the origin of the dark GRBs is dust obscuration along the line of sight to GRBs \citep[e.g.,][]{perl09}. 
It has been found that dust-obscured GRB host galaxies are more massive than the those of optically bright GRBs \citep[e.g.,][]{kruh11, perl13}. 
\cite{whit17} found a dependence of the fraction of obscured star formation (${\rm SFR_{IR}/SFR_{UV+IR}}$) on stellar mass out to $z = 2.5$: 50\% of star formation is obscured for galaxies with $\log(M/M_{\odot}) = 9.4$, and $>$90\% for galaxies with $\log(M/M_{\odot}) > 10.5$. 
Therefore, it is natural for dark GRBs to be hosted by relatively massive galaxies. 
GRB~080207 is identified as a dark GRB, and the host galaxy is a massive and high-metallicity galaxy. 
We found in this study that the host galaxy shares common properties in terms of molecular gas, such as gas mass fraction, gas-to-dust ratio, location in the $M_{\rm gas}$--SFR or $\Sigma_{\rm gas}$--$\Sigma_{\rm SFR}$ relation, and gas kinematics, with normal MS galaxies, which dominate the cosmic star formation activity, suggesting that can GRBs occur in normal star-forming environments at $z \sim 2$. 
Currently the sample of GRB host galaxies with CO detection is limited and only one host galaxy is detected at $z > 1$. 
It is essential to increase the sample size in order to examine the general properties of molecular gas in GRB host galaxies and the availability of GRBs as an unbiased tracer of cosmic SFR density. 
ALMA and NOEMA are now routinely detecting molecular gas in MS star-forming galaxies at $z > 1$. 
As discussed in this paper, in order to study CO line ratios and molecular gas excitation conditions in GRB host galaxies, observations of ground state CO transition are also required by using longer-wavelength sensitive facilities such as VLA.

\section{Conclusions}\label{sec:conclusions}

We present the results of CO(1--0) and CO(4--3) observations of the GRB~080207 host galaxy at $z = 2.0858$ by using VLA and ALMA. 
The CO(1--0) and CO(4--3) lines are detected, making the host galaxy the first case in GRB host galaxies with more than two CO transitions together with the CO(2--1) and CO(3--2) lines in the literature. 
The CO(4--3) data is spatially resolved, allowing us to estimate the kinematics and morphological parameters of the host galaxy. 
We derived physical quantities based on the observations:

By using the CO(1--0) line luminosity and adopting a metallicity-dependent conversion factor of $\alpha_{\rm CO} = 5.1$ $M_{\odot}$~(K~km~s$^{-1}$~pc$^2$)$^{-1}$, we derive a molecular gas mass of $M_{\rm gas} = 8.7 \times 10^{10}$~$M_{\odot}$. 
We found that host galaxy is located in the similar region for $z \sim 1$--2 MS galaxies and on the sequence of normal star-forming galaxies in the $M_{\rm gas}$--SFR plane. 
The molecular gas mass fraction is $M_{\rm gas}/(M_{\rm gas} + M_*) = 0.34$, which is comparable to $z \sim 1$--2 MS galaxies with the similar stellar mass.

We fit a modified black body to the FRI--millimeter photometry, and derived a dust temperature of $T_{\rm dust} = 37 \pm 4$~K, emissivity index of $\beta = 1.5 \pm 0.3$, and a dust mass of $M_{\rm dust} = 1.5 \times 10^8$~$M_{\odot}$. 
The molecular gas-to-dust mass ratio of $M_{\rm gas}/M_{\rm dust} \sim 560$ is consistent with $z \sim 1.4$ MS galaxies with the similar stellar mass and metallicity.

The model fitting to the spatially resolved CO(4--3) cube data shows that the host galaxy is a rotation-dominated disk with a CO half-light radius of $Rr_{\rm 1/2,CO} = 2.4$~kpc and a rotation velocity of $v_{\rm rot} = 348$~km~s$^{-1}$. 
The ratio of rotational motion to velocity dispersion of $v_{\rm rot}/\sigma_0 = 5.5$ is comparable to those of $z \sim 1$--2 MS galaxies. 
The dynamical mass is comparable to the sum of the molecular gas mass and stellar mass, suggesting that the contribution of dark matter is not significant.

The surface densities of molecular gas and SFR show that the host galaxy follows the relation of normal star-forming galaxies in the $\Sigma_{\rm gas}$--$\Sigma_{\rm SFR}$ plane. 
The CO spectral line energy distribution derived from the four CO transitions is similar to that of starburst galaxies, suggesting a high excitation condition of molecular gas. 
If we introduce the dynamical timescale, the host galaxy is found to be on the general star-formation law for local and high-$z$ galaxies in the $\Sigma_{\rm gas}/\tau_{\rm dyn}$--$\Sigma_{\rm SFR}$ plane.

The CO line ratios show that molecular gas excitation is high and close to unity up to $J = 4$, suggesting the presence of dense and warm gas. 
The CO SLED is consistent with the model of high $\Sigma_{\rm SFR}$.

We found that the host galaxy and normal MS galaxies at similar redshifts share common properties, suggesting that long-duration GRBs can occur in normal star-forming environments at $z \sim 2$.

\acknowledgments

We would like to acknowledge Drew Medlin, Heidi Medlin, staffs at NRAO and ALMA Regional Center for their help in preparation for VLA and ALMA observations and data acquisition. 
We thank the referee for helpful comments and suggestions. 
We are grateful to Ken-ichi Tadaki for useful discussions. 
BH is supported by JSPS KAKENHI Grant Number 15K17616. 
LVT was supportted by the OTKA grant NN-111016. 
This study is supported by the ALMA Japan Research Grant of NAOJ Chile Observatory, NAOJ-ALMA-209. 
The National Radio Astronomy Observatory is a facility of the National Science Foundation operated under cooperative agreement by Associated Universities, Inc.
This paper makes use of the following ALMA data: ADS/JAO.ALMA\#2015.1.00939.S. 
ALMA is a partnership of ESO (representing its member states), NSF (USA) and NINS (Japan), together with NRC (Canada), MOST and ASIAA (Taiwan), and KASI (Republic of Korea), in cooperation with the Republic of Chile. The Joint ALMA Observatory is operated by ESO, AUI/NRAO and NAOJ. 
Based on observations made with the NASA/ESA Hubble Space Telescope, and obtained from the Hubble Legacy Archive, which is a collaboration between the Space Telescope Science Institute (STScI/NASA), the Space Telescope European Coordinating Facility (ST-ECF/ESA) and the Canadian Astronomy Data Centre (CADC/NRC/CSA).

{\it Facilities:} \facility{VLA, ALMA}. 


\end{document}